\documentclass[aps,twocolumn,showlabels,showrefs,amsmath,amssymb,prl,superscriptaddress,floatfix,colors]{revtex4-1}

\usepackage{graphicx}
\usepackage{dcolumn}
\usepackage{bm}
\usepackage{amssymb}
\usepackage{multirow}
\usepackage{color}
\usepackage{braket}
\usepackage[normalem]{ulem}
\usepackage{mathtools}

\usepackage{xcolor}
\definecolor{blueprl}{RGB}{13.0, 18.0, 180.0 }
\usepackage[colorlinks=true, allcolors=blueprl]{hyperref}

\newcommand{\Pe}{\mathrm{Pe}}

\begin{document}

\title{Work fluctuations for a harmonically confined Active Ornstein-Uhlenbeck Particle}

\date{\today}
\author{Massimiliano Semeraro}
\affiliation{Dipartimento  Interateneo di  Fisica,  Universit\`a  degli  Studi  di  Bari  and  INFN,
Sezione  di  Bari,  via  Amendola  173,  Bari,  I-70126,  Italy}

\author{Giuseppe Gonnella}
\affiliation{Dipartimento Interateneo  di  Fisica,  Universit\`a  degli  Studi  di  Bari  and  INFN,
Sezione  di  Bari,  via  Amendola  173,  Bari,  I-70126,  Italy}

\author{Antonio Suma}
\affiliation{Dipartimento Interateneo  di  Fisica,  Universit\`a  degli  Studi  di  Bari  and  INFN,
Sezione  di  Bari,  via  Amendola  173,  Bari,  I-70126,  Italy}

\author{Marco Zamparo}
\affiliation{Dipartimento  Interateneo di  Fisica,  Universit\`a  degli  Studi  di  Bari  and  INFN,
Sezione  di  Bari,  via  Amendola  173,  Bari,  I-70126,  Italy}

\email[email: ]{name@}

\begin{abstract}
  We study the active work fluctuations of an active
  Ornstein-Uhlenbeck particle in the presence of a confining harmonic
  potential. We tackle the problem analytically both for stationary
  and generic uncorrelated initial states. Our results show that
  harmonic confinement can induce singularities in the active work
  rate function, with linear stretches at large positive and negative
  active work, at sufficiently large active and harmonic force
  constants. These singularities originate from big jumps in the
  displacement and in the active force, occurring at the initial or
  ending points of trajectories and marking the relevance of boundary
  terms in this problem.
\end{abstract}

\maketitle

Large deviation theory has a profound impact in statistical physics
\cite{ellis2012entropy, touchette2009b}. In non-equilibrium systems,
where probability measures on configuration spaces are not naturally available,
it provides an analogous of the usual equilibrium free-energy
description. Given an extensive physical observable $W_{\tau}$
computed by cumulating a large number $\tau$ of microscopic events, if
a large deviation principle holds, then the asymptotics of the
probability distribution $P(W_\tau/\tau=w)$ can be characterized by
the rate function $I(w)=-\lim_{\tau\uparrow\infty}\frac{1}{\tau}\ln
P(W_\tau/\tau=w)$ \cite{dembo1984, denhollander2000}. The probability
distribution is dominated by small fluctuations around the minimum of
$I$, which in this sense plays a role similar to a free-energy.

Singularities in  rate functions can be seen as
the hallmarks of  phase transitions \cite{touchette2009b, jack2020}. They appear in different contexts, 
such as in studies of heat
exchanges, diffusive transport, and entropy
production \cite{bodineau2005, harris2005, visco2006,PhysRevE.78.011123_2008,
  hurtado2011, bertini2010, lefevere2011,bunin2012, Speck_2012,
  PhysRevE.90.042123_2014, Tsobgni_Nyawo_2016, harris2017,
  zamparo2019}, and in some cases have been interpreted
as due to a condensation mechanism
\cite{jeon2000,majumdar2005,ARMENDARIZ2009,merhav2010,nossan2014, corberi2013,Zannetti_2015,godrche2019,zamparojpa2022}. If $\tau$ is
a time interval, the rate function can provide a generalized
thermodynamic description based on the counting of trajectories, and
the singularity would correspond to a phase separation in trajectory space
\cite{garrahan2009, jack2010,tailleur2007}.

Active matter systems \cite{gompper2020}, with their inherent nonequilibrium character, offer
a new field for applications of large deviation theory
and investigations on dynamical phase transitions. In these systems, available energy
sources are locally employed to produce spontaneous motion or work on the environment.
They are characterized  by a  surprisingly rich phenomenology, including  
 new important  phenomena like motility induced phase
separation (MIPS) \cite{tailleur2008} or spontaneous flow \cite{gompper2020}, and also concerning fluctuation properties
  \cite{pietzonka2016, cagnetta2017, whitelam2018,
  grandpre2018, gradenigo2019, cagnetta2020,
  fodor2020, Chaudhuri2021,grandpre2021,smith2022}.

A crucial quantity for the description of dynamical transitions in
active matter models is the active work, defined as the time-average
of the power of the propulsion force. In dilute systems of active
Brownian particles, the active work rate function was shown to be
singular \cite{cagnetta2017}, with a linear tail associated to
trajectories where a particle does not diffuse freely but is dragged
by a cluster moving oppositely to its propulsion force.
Successive studies have revealed a very rich structure for the phase diagram
in trajectory space \cite{nemoto2019, keta2021, agranov2022}.
Rate functions for quantities analogous to active work were investigated in experiments of
polar beads embedded in two-dimensional granular layers
\cite{kumar2011} and idealized Maxwell-Lorentz granular
systems \cite{gradenigo2013}.

Most of the aforementioned results were based on
numerical, although sophisticated, studies of interacting particle
models. Rigorous analysis of simpler models can help to elucidate  
the role of self-propulsion in dynamical transitions. 
In this Letter we consider a single active Ornstein-Uhlenbeck particle
(AOUP) \cite{szamel2014, maggi2015, farage2015,
fodor2016, caprini2019, martin2021,
crisanti2022,caraglio2022,arsha2022, szamel2022} 
and investigate analytically the active work fluctuations in the presence of a confining harmonic
potential. AOUP systems share many of the relevant properties of other
interacting active particle models, including MIPS. Restricting to one
particle description, a confining potential can mimic the
trapping created by other particles at finite densities
\cite{szamel2014,nandi2017,woillez2020}. 
We will show that, differently from the case of
a free AOUP \cite{semeraro2021}, harmonic confinement can induce
singularities in the active work rate function, with linear
stretches at large positive and negative active work. These
singularities are found both for stationary and generic
uncorrelated initial states at sufficiently large active and
harmonic force constants.  They originate from big jumps in the
displacement and in the active force, occurring at the initial or
ending points of trajectories and marking the relevance of boundary
terms in this problem.  

The unidimensional active particle model that we study is
defined 
via the Ornstein-Uhlenbeck process
\begin{equation} 
\begin{cases}
	\gamma\dot{r}(t)=a(t)-k r(t)+\sqrt{2\gamma k_BT}\,\xi(t) &\\
	 \dot{a}(t) =-\nu a(t)+ F\sqrt{2\nu}\,\eta(t),
	 \end{cases}
\label{eq:aoup_model}
\end{equation}
where $r(t)$ is the position of a unit-mass particle in a harmonic
potential of elastic constant $k$, $a(t)$ represents a self-propulsion
force with amplitude $F$ and decay rate $\nu$, $\gamma$ and $T$ are
the friction coefficient and the bath temperature, and $\xi(t)$ and
$\eta(t)$ are two independent zero-mean and unit-variance white
noises. One has $\nu\equiv k_BT/\gamma d^2$ with $d$ being a length
proportional to the particle's diameter and related to the ratio
between the rotational and translational diffusion coefficients
\cite{das2018}. We will vary the adimensional elastic constant
$\kappa\equiv\frac{kd^2}{k_BT}$ and the \textit{P\'eclet number}
$\Pe\equiv\frac{Fd}{k_BT}$, which quantify the strength of the
potential and of the active force with respect to thermal
fluctuations \cite{das2018, mandal2019}. 
Without loss of generality, in the following we set $\gamma=1$, $k_BT=1$, and $d=1$.

In order to analyze the dynamical behavior of the AOUP model
(\ref{eq:aoup_model}), we examine the probability distribution of the
{\it active work} $W_\tau$ defined by the formula \footnote{T{he}
  results we present here are based on the Stratonovich
  prescription. We have checked that the It\^o definition leads to the
  same findings}
\begin{equation*}
W_\tau\equiv\int_0^\tau a(t)\dot{r}(t)\,dt.
\end{equation*} 
The active work represents a measure of how efficiently
self-propulsion is converted into directed motion. Our goal is to
evaluate the rate function
$I(w)=-\lim_{\tau\uparrow\infty}\frac{1}{\tau}\ln P (W_\tau/\tau=w)$.
The probability distribution $P(W_\tau/\tau=w)$ can be expressed by
the path integral
\begin{align}
  \nonumber
P (W_\tau/\tau=w )&=\int \delta(W_\tau-w\tau)\,\mathcal{P}_\tau\,\mathcal{D} r\mathcal{D} a
\end{align}
with path probability
\begin{align}
  \nonumber
 & \mathcal{P}_\tau \propto
  \exp\bigg\{\!-\frac{1}{2}\begin{psmallmatrix}r(0) & a(0)\end{psmallmatrix}\Sigma_0^{-1}\begin{psmallmatrix} r(0) \\ a(0)\end{psmallmatrix}\bigg\}\times\\
  \nonumber
&\exp\bigg\{\!-\!\!\int_0^\tau \!\!\bigg( \frac{[\dot{r}(t)-a(t)+\kappa\, r(t)]^2}{4}+\frac{[\dot{a}(t)+a(t)]^2}{4\Pe^2}\bigg)dt\bigg\}.
\nonumber
\end{align}
The probability $\mathcal{P}_\tau$ combines the distribution of the
initial values $r(0)$ and $a(0)$ with the Onsager-Machlup weight for
the trajectory \cite{onsager1953I}.  We have chosen
Gaussian initial data with mean zero and joint covariance matrix
$\Sigma_0$. In particular, we focus on a non-stationary uncorrelated
initial condition with standard deviations $\sigma_r$ for $r(0)$ and
$\sigma_a$ for $a(0)$, and on the stationary case given by
\cite{Gardiner2003}
\begin{equation}
  \Sigma_0=\begin{pmatrix}
  \frac{1+\kappa+\Pe^2}{\kappa(1+\kappa)} & \frac{\Pe^2}{1+\kappa} \\
  \frac{\Pe^2}{1+\kappa} & \scalebox{0.73}{$\Pe^2$}
  \end{pmatrix}.
  \label{eq:staz_incond}
\end{equation}

An operative definition of the rate function $I$ requires to first
look at a discrete-time problem with time step $\epsilon$, and then to
consider the continuum limit $\epsilon\downarrow 0$. In fact, we
compute $I$ by means of the double limit
$I(w)=-\lim_{\epsilon\downarrow
  0}\lim_{N\uparrow\infty}\frac{1}{N\epsilon}\ln P (W_N/N\epsilon =w
)$, where $W_N\equiv\frac{1}{2}\sum_{n=1}^N(a_n+a_{n-1})(r_n-r_{n-1})$
with $r_n\equiv r(n\epsilon)$ and $a_n\equiv a(n\epsilon)$ is the
discretized active work at time $N\epsilon$.
The discrete-time problem is tackled by computing the asymptotic
cumulant generating function $\frac{1}{N}\ln\Braket{e^{\lambda W_N}}$
of $W_N$ at large $N$. The Legendre-Fenchel transform of
$\lim_{N\uparrow\infty}\frac{1}{N}\ln\Braket{e^{\lambda W_N}}$ with
respect to the additional variable $\lambda$ is expected to be the
discrete-time rate function
$J(w)=-\lim_{N\uparrow\infty}\frac{1}{N}\ln P (W_N/N=w)$. We have
$I(w)=\lim_{\epsilon\downarrow 0}\frac{J(\epsilon w)}{\epsilon}$.

At small $\epsilon$, the trajectory $\{(r_0,a_0),\ldots,(r_N,a_N)\}$
is distributed according to a multivariate Gaussian law with mean zero
and covariance matrix $\Sigma_N$ \footnote{See Supplemental
  Material for additional information}.  Regarding $W_N$ as a
quadratic functional of $\{(r_0,a_0),\ldots,(r_N,a_N)\}$ with
coefficient matrix $\frac{1}{2}\mathsf{M}_N$, a standard Gaussian
integral gives
\begin{align}
  \nonumber
  \ln\Braket{e^{\lambda W_N}}&=-\frac{1}{2}\ln\det(\Sigma_N^{-1}-\lambda \mathsf{M}_N)\\
\nonumber
  &-N\ln(2\epsilon\Pe)-\frac{1}{2}\ln\det\Sigma_0
\end{align}
if $\Sigma_N^{-1}-\lambda \mathsf{M}_N$ is positive definite and
$\ln\Braket{e^{\lambda W_N}}=+\infty$ otherwise.
$\Sigma_N^{-1}-\lambda \mathsf{M}_N$ is the block tridiagonal matrix
\begin{equation}
  \Sigma_N^{-1}-\lambda \mathsf{M}_N=\begin{pmatrix}
  L & V^\top &  & & \\
    V & U & \ddots &  &\\
      & \ddots & \ddots & \ddots  & \\
     & & \ddots & U & V^\top\\
     & &  & V & R
  \end{pmatrix}
  \label{matriciona}
\end{equation}
with $2\times 2$ blocks $L\equiv\Sigma_0^{-1}+S^\top D^{-2}S+\lambda
E_+$, $U\equiv D^{-2}+S^\top D^{-2}S$, $R\equiv D^{-2}-\lambda E_+$,
$V\equiv-D^{-2}S-\lambda E_-$,
$S\equiv\begin{psmallmatrix}1-\kappa\epsilon & \epsilon\\ 0 &
1-\epsilon\end{psmallmatrix}$,
$D\equiv\begin{psmallmatrix}\sqrt{2\epsilon} & 0\\ 0 &
\Pe\sqrt{2\epsilon}\end{psmallmatrix}$, and
$E_\pm\equiv\frac{1}{2}\begin{psmallmatrix}0 & 1\\ \pm 1 &
  0\end{psmallmatrix}$. We observe that $\Sigma_N^{-1}-\lambda
  \mathsf{M}_N$ differs from a perfect block tridiagonal Toeplitz
  matrix by the extreme diagonal blocks $L$ containing $\Sigma_0$ and
  $R$, which play a subtle but important role in determining positive
  definiteness.  We denote by $\mathsf{T}_N$ the block Toeplitz bulk
  matrix obtained from $\Sigma_N^{-1}-\lambda \mathsf{M}_N$ by
  deleting all contour blocks.

Evaluating the limit
$\lim_{N\uparrow\infty}\frac{1}{N}\ln\Braket{e^{\lambda W_N}}$ is a
nontrivial task. For those values of $\lambda$ that make
$\Sigma_N^{-1}-\lambda \mathsf{M}_N$ positive definite, the asymptotic
cumulant generating function of $W_N$ is only determined by the bulk
matrix $\mathsf{T}_N$. In fact, the results of \cite{zamparo2022} for
generic quadratic functionals based on Szeg\"o theorem for block
Toeplitz matrices \cite{gutierrez2008} show that if
$\Sigma_N^{-1}-\lambda \mathsf{M}_N$ is positive definite in the large
$N$ limit, then
\begin{align}
  \nonumber
  \lim_{N\uparrow\infty}\frac{1}{N}\ln\Braket{e^{\lambda W_N}}&=\varphi(\lambda)\\
  \nonumber
  &\equiv-\frac{1}{4\pi}\int_0^{2\pi}\ln\det F_{\lambda}(\theta)\,d\theta-\ln(2\epsilon\Pe)
\end{align}
with $F_\lambda(\theta)\equiv V e^{-\mathrm{i}\theta}+U+V^\top
e^{\mathrm{i}\theta}$. The Hermitian matrix function $F_\lambda$ is
the so-called \textit{symbol} of $\mathsf{T}_N$ \cite{gutierrez2008}.

For $\Sigma_N^{-1}-\lambda \mathsf{M}_N$ being positive definite it is
necessary and sufficient that both $\mathsf{T}_N$ and its Schur
complement
\begin{equation*}
  \mathsf{S}_N\equiv\begin{pmatrix}
    L-V^\top(\mathsf{T}_N^{-1})_{11}V & -V^\top(\mathsf{T}_N^{-1})_{1N}V^\top\\
    -V(\mathsf{T}_N^{-1})_{N1}V & R-V(\mathsf{T}_N^{-1})_{NN}V^\top
  \end{pmatrix}
\end{equation*}
are positive definite, $(\mathsf{T}_N^{-1})_{ij}$ being the $2\times
2$ block of $\mathsf{T}_N^{-1}$ in the row $i$ and column $j$. The
extreme diagonal blocks $L$ and $R$ enter $\mathsf{S}_N$ and thus come
into play in establishing positive definiteness. The Toeplitz matrix
$\mathsf{T}_N$ is positive definite if its symbol $F_\lambda$ has the
same property \cite{zamparo2022}. This introduces a first constraint
on $\lambda$, which defines the \textit{primary domain} $(\tilde
l_-,\tilde l_+)$ of $\varphi$. It can be shown \cite{zamparo2022} that
$\lim_{N\uparrow\infty}\mathsf{S}_N=\begin{psmallmatrix}\mathcal{L}_\lambda
& 0\\ 0 & \mathcal{R}_\lambda\end{psmallmatrix}$,
$\mathcal{L}_\lambda$ and $\mathcal{R}_\lambda$ being $2\times2$
symmetric matrices determined by $L$, $R$, and $F_\lambda$, whose
explicit expression in the limit $\epsilon\downarrow 0$ is reported in
\cite{Note2}. Then, a second constraint on $\lambda$ comes from the
requirement that $\mathcal{L}_\lambda$ and $\mathcal{R}_\lambda$ are
positive definite. Denoting by $(l_-,l_+)$ the interval of $\lambda$ for which both
constraints are fulfilled, i.e.\ the \textit{effective domain} of
$\varphi$, we get
$\lim_{N\uparrow\infty}\frac{1}{N}\ln\Braket{e^{\lambda
    W_N}}=\varphi(\lambda)$ for $\lambda\in(l_-,l_+)$ and
$\lim_{N\uparrow\infty}\frac{1}{N}\ln\Braket{e^{\lambda W_N}}=+\infty$
for $\lambda\notin[l_-,l_+]$. We have $\tilde{l}_-\le l_-<0<l_+\le
\tilde{l}_+$.

We are now in the position to compute the discrete-time rate function
$J$ as the Legendre-Fenchel transform of $\varphi$, that is
$J(w)=\sup_{\lambda\in(l_-,l_+)}\{w\lambda-\varphi(\lambda)\}$.  We
stress that this way of computing the rate function $J$, although
natural, cannot be justified by the classical G\"artner-Ellis theorem
\cite{dembo1984, denhollander2000} since in general $\varphi$ is not
steep at the boundary of the effective domain. In fact, the
G\"artner-Ellis theorem requires that the derivative of $\varphi$
diverges when the boundary points $l_-$ and $l_+$ are approached, but
this fails when $l_->\tilde{l}_-$ or $l_+<\tilde{l}_+$.  The above
formula for $J$ can be demonstrated by means of a time-dependent
change of probability measure \cite{zamparo2022}. From a mathematical
point of view, the lack of steepness is the hallmark of a dynamical
phase transition.

The last job is to take the continuum limit. Notice that $\tilde
l_\pm$, $l_\pm$, and $\varphi(\lambda)$ depend on
$\epsilon$. Cumbersome calculations summarized in \cite{Note2} yield
\begin{equation}
  I(w)=\lim_{\epsilon\downarrow 0}\frac{J(\epsilon w)}{\epsilon}=\sup_{\lambda\in(\lambda_-,\lambda_+)}\big\{w\lambda-\phi(\lambda)\big\}
\label{I_final}
\end{equation}
with $\lambda_\pm\equiv\lim_{\epsilon\downarrow 0}l_\pm$ and
asymptotic cumulant generating function
\begin{align}
  \nonumber
  \phi(\lambda)&\equiv\lim_{\epsilon\downarrow 0}\frac{\varphi(\lambda)}{\epsilon}\\
  &=\frac{1+\kappa}{2}-\frac{1}{2}\sqrt{(1+\kappa)^2-4\Pe^2\lambda(1+\lambda)}.
  \label{def:phi}
\end{align}
The primary domain in the limit $\epsilon\downarrow 0$ is found to be
described by the compact formula
\begin{equation}
  \tilde\lambda_\pm\equiv\lim_{\epsilon\downarrow 0}\tilde l_\pm=-\frac{1}{2}\pm\frac{1}{2}\sqrt{1+\bigg(\frac{1+\kappa}{\Pe}\bigg)^2}.
  \label{eq:primary_bounds}
\end{equation}
An explicit formula for the boundary points $\lambda_\pm$ of the
effective domain in the continuum limit is not available.

According to Eq.\ (\ref{def:phi}), the asymptotic cumulant generating
function $\phi$ is steep on the primary domain
$(\tilde\lambda_-,\tilde\lambda_+)$ as
$\lim_{\lambda\downarrow\tilde\lambda_-}\phi'(\lambda)=-\infty$ and
$\lim_{\lambda\uparrow\tilde\lambda_+}\phi'(\lambda)=+\infty$. On the
contrary, $w_-\equiv\phi'(\lambda_-)>-\infty$ if
$\lambda_->\tilde\lambda_-$ and $w_+\equiv\phi'(\lambda_-)<+\infty$ if
$\lambda_+<\tilde\lambda_-$, so that $\phi$ is not steep on the
effective domain $(\lambda_-,\lambda_+)$ when
$\lambda_->\tilde\lambda_-$ or $\lambda_+<\tilde\lambda_+$. The lack
of steepness originates linear tails of the rate function $I$ that
begin at the singular points $w_-$ and $w_+$. In fact, the supremum in
Eq.\ (\ref{I_final}) reads
\begin{equation*}
  I(w)=\begin{cases}
  \lambda_-(w-w_-)+i(w_-) & \mbox{if }w\le w_-,\\
  i(w) &\mbox{if }w_-<w<w_+,\\
  \lambda_+(w-w_+)-i(w_+)  & \mbox{if }w\ge w_+
  \end{cases}
\end{equation*}
with
$i(w)\equiv\frac{\sqrt{1+(w/\Pe)^2}\sqrt{(1+\kappa)^2+\Pe^2}-1-\kappa-w}{2}$.
Interestingly, the smooth function $i$ is the rate function of the entropy production at
stationarity \cite{Note2}, and as such satisfies the Gallavotti-Cohen
symmetry $i(-w)=i(w)+w$ at variance with $I$. The entropy
production differs from the active work by local contributions of the
initial and ending points of the trajectory \cite{Note2}, which
prevent its rate function from exhibiting singularities at
stationarity \cite{Note2}, a circumstance
that boosts the interest in the active work.

\begin{figure}
\centering 
\includegraphics[width=8.6cm]{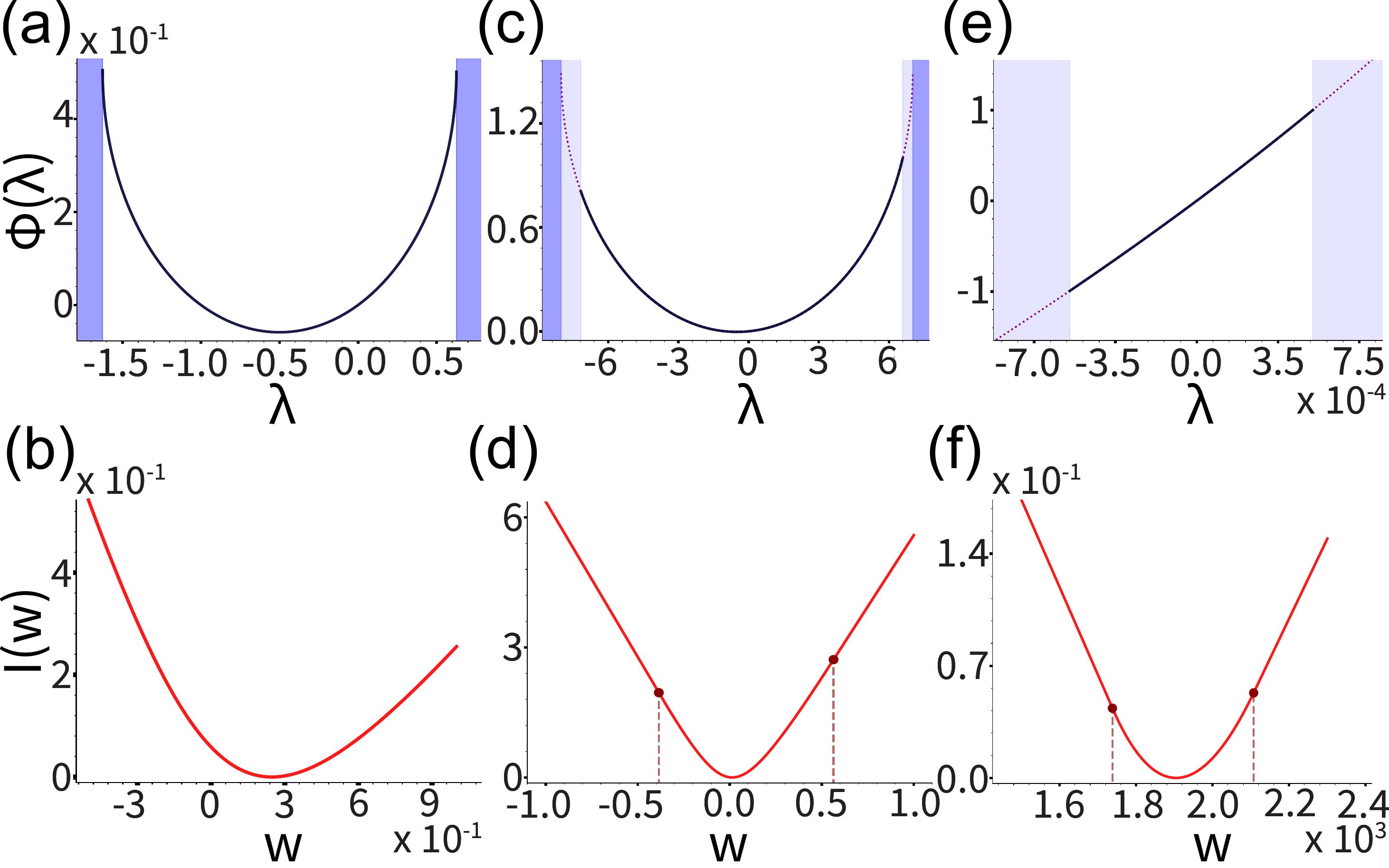}
\vspace{-0.25cm}
\caption{\footnotesize{Asymptotic cumulant generating function
    $\phi(\lambda)$ and rate function $I(w)$ under a concentrated
    non-stationary initial condition with $\sigma_r\downarrow 0$ and
    $\sigma_a\downarrow 0$ for $\kappa=0.01$ and $\Pe=0.5$ in (a) and
    (b), under the stationary initial condition with $\kappa=2.0$ and
    $\Pe=0.2$ in (c) and (d), and under the stationary initial
    condition with $\kappa=20.0$ and $\Pe=200.0$ in (e) and (f). The
    dark and light blue areas in (a), (c), and (e) mark the regions
    outside the primary and effective domain, respectively. The dotted
    lines in (d) and (f) mark the beginning of the left linear tail at
    $w_-$ and of the right one at $w_+$.}}
\label{fig:fig1}
\end{figure}

\begin{figure}
  \centering
  \includegraphics[width=8.0cm,height=7cm]{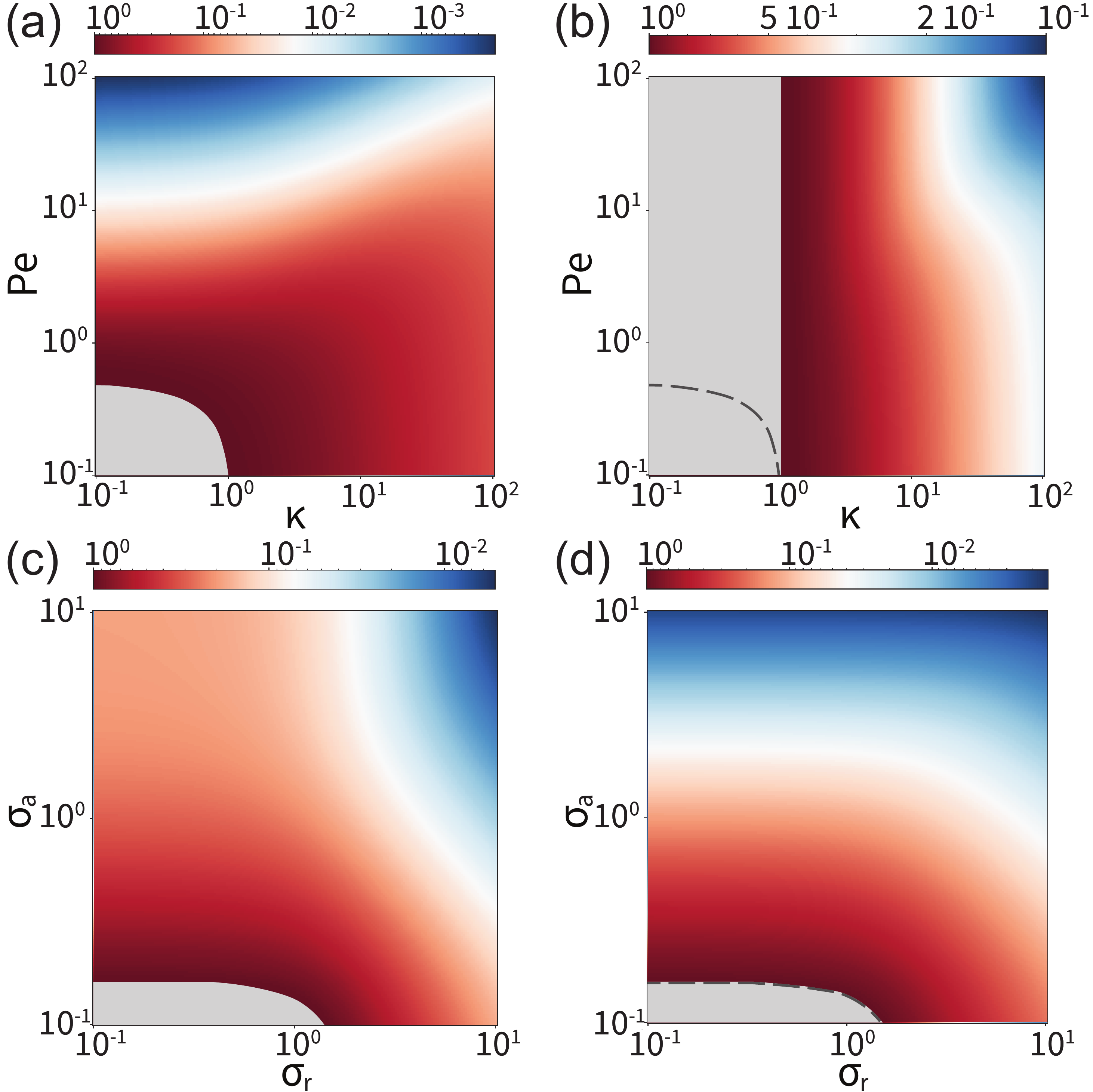}
\vspace{-0.25cm}
\caption{\footnotesize{Phase diagram as deduced by the ratios $r_-\equiv\lambda_-/\tilde{\lambda}_-$ and $r_+\equiv\lambda_+/\tilde{\lambda}_+$
    between the effective and primary domain boundary points
    of the asymptotic cumulant generating function.
    Colored areas denote regions where a dynamical phase transition
    occurs, i.e.\ $r_-<1$ or $r_+<1$, and the color scale measures
    $r_-$ and $r_+$.  Gray areas denote regions without a singularity,
    i.e.\ $r_-=1$ or $r_+=1$. (a) and (b): $r_-$ and $r_+$ under the
    stationary initial condition in the $\kappa-\Pe$ plane.  (c) and
    (d): $r_-$ and $r_+$ under the non-stationary initial condition in the
    $\sigma_r-\sigma_a$ plane at 
    $\kappa=0.7$ and $\Pe=0.1$ for which
    there is no dynamical phase transition at stationarity. The regions
    under the dashed lines do not exhibit any phase transition.}}
\label{fig:fig2}
\end{figure}

Fig.\ \ref{fig:fig1} shows the functions $\phi$ and $I$.
Figs.\ \ref{fig:fig1}(a) and \ref{fig:fig1}(b) refer to the
concentrated non-stationary initial condition $\sigma_r\downarrow 0$
and $\sigma_a\downarrow 0$, for which the primary and the effective
domain coincide. Figs.\ \ref{fig:fig1}(c) and \ref{fig:fig1}(d), and
Figs.\ \ref{fig:fig1}(e) and \ref{fig:fig1}(f), correspond to
stationary initial conditions with different parameters. At
stationarity the rate function has a left linear tail,
i.e.\ $\lambda_->\tilde\lambda_-$, for
$\Pe\sqrt{(3+\kappa)(1+3\kappa)}>1-\kappa^2$ and a right linear tail,
i.e.\ $\lambda_+<\tilde\lambda_+$, for $\kappa>1$
\cite{Note2}. Fig.\ \ref{fig:fig2} reports the phase diagrams of the
system as deduced by inspecting the ratios
$r_-\equiv\lambda_-/\tilde{\lambda}_-$ and
$r_+\equiv\lambda_+/\tilde{\lambda}_+$. Figs.\ \ref{fig:fig2}(a) and
\ref{fig:fig2}(b) show that at stationarity and at large $\kappa$ and
$\Pe$ the effective domain is significantly smaller than the primary
domain with $\lambda_-\gg\tilde{\lambda}_-$ or
$\lambda_+\ll\tilde{\lambda}_+$, respectively.
Fig.\ \ref{fig:fig2}(c) and Fig.\ \ref{fig:fig2}(d) depict $r_-$ and
$r_+$ under non-stationary initial conditions for fixed values of
$\kappa$ and $\Pe$ such that the corresponding stationary problem has
no linear tail.  The effective domain is significantly smaller than
the primary domain with $\lambda_-\gg\tilde{\lambda}_-$ or
$\lambda_+\ll\tilde{\lambda}_+$ at large $\sigma_r$ and $\sigma_a$.
We note that the results of Ref.\ \cite{semeraro2021} on the free AOUP
are consistently recovered in the limit $\kappa\downarrow 0$ by the
confined non-stationary model with $\sigma_a=\Pe$
\cite{Note2}. Singularities of the rate function are lost in this
limit.

\begin{figure}
\centering 
\includegraphics[width=9.0cm]{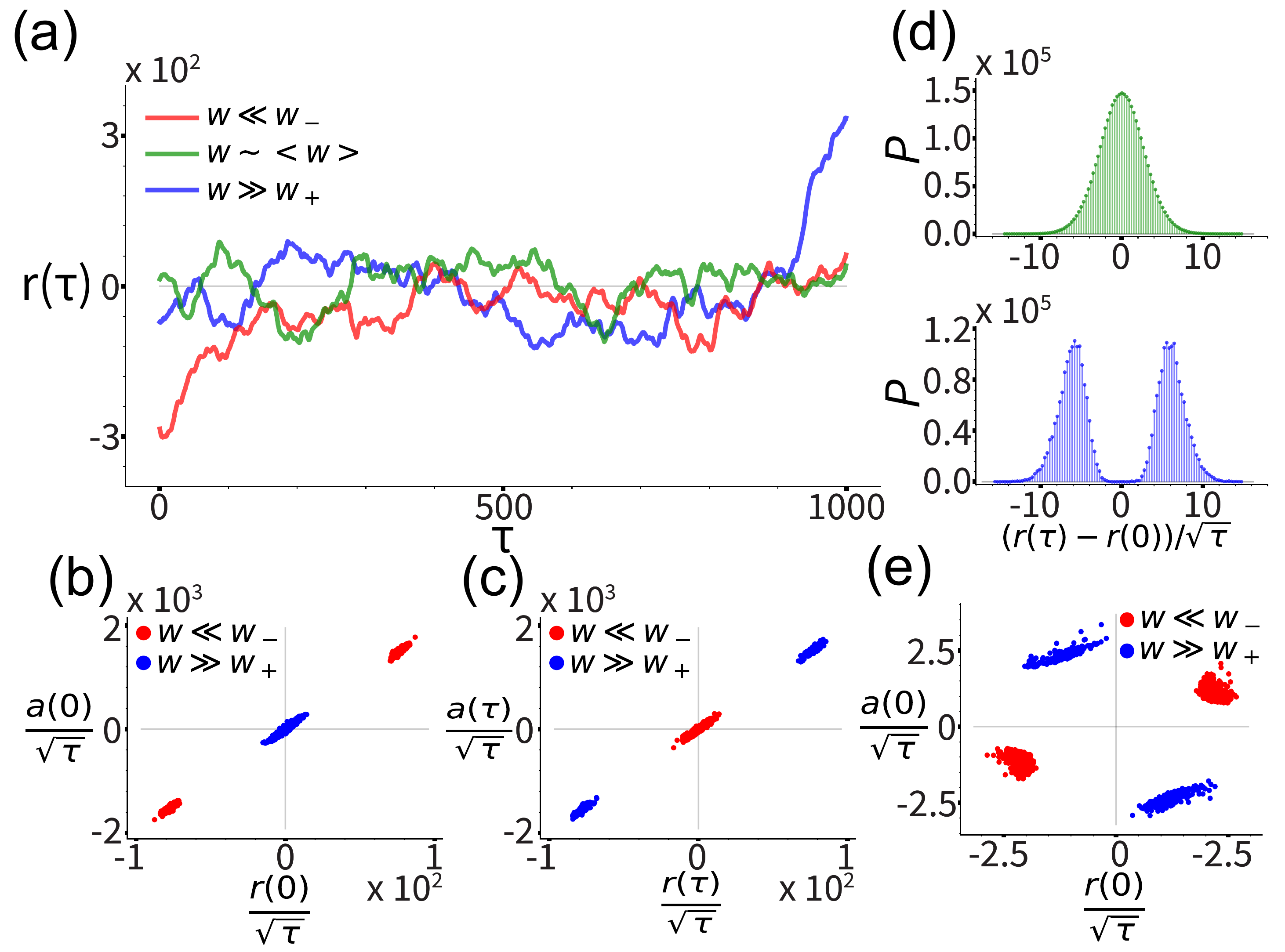}
\vspace{-0.25cm}
\caption{\footnotesize{Trajectory analysis at stationarity with
    $\kappa=20.0$ and $\Pe=200.0$ and
    under the non-stationary initial condition with $\kappa=0.7$ and
    $\Pe=0.1$.  (a): typical trajectories of the particle in the
    stationary configuration up to time $\tau=10^3$ corresponding to
    $w=26\ll w_-=1.73\cdot 10^3$ (red), $w=1.92\cdot
    10^3\approx\braket{w}$ (green), and $w=4.10\cdot10^3\gg
    w_+=2.11\cdot10^3$ (blue). (b) and (c): initial and ending
    points of a pool of stationary trajectories corresponding to
    $w\leq 8.00\cdot 10^2\ll w_-$ and $w\geq3.20\cdot 10^3\gg w_+$,
    respectively. (d): stationary probability distribution of the
    particle's net displacement conditional on typical $w$ in the
    interval $-\sigma_w<w-\braket{w}<\sigma_w$ (top) and on large $w$
    in the interval $3\sigma_w<w-\braket{w}<4\sigma_w$ (bottom),
    $\sigma_w$ being the standard deviation of the active work taking
    value $\sigma_w\sim2.76\cdot 10^2$ at $\tau=10^3$.  (e): initial
    and ending points of a pool of non-stationary trajectories with 
$\sigma_r=\sigma_a=10$ and $\tau=2\cdot10^4$ corresponding to $w\leq
    -5.00\cdot 10^{-2}\ll w_-=1.09\cdot 10^{-3}$ and $w\geq2.50\cdot
    10^{-1}\gg w_+=8.19\cdot 10^{-2}$, respectively.}}
\label{fig:fig3}
\end{figure}

Interpretation of the singularities of the rate function requires to
analyze the particle trajectories.  Fig.\ \ref{fig:fig3}(a) reports
three typical trajectories at stationarity with large $\kappa$ and
$\Pe$ conditional on $W_\tau=w\tau$ with $w\ll w_-$ in the far left
linear tail, $w\approx\braket{w}$, and $w\gg w_+$ in the far right
linear tail. $\braket{w}$ is the typical value of the active work,
that is $I(\braket{w})=0$. A large fluctuation $w\ll w_-$ of the
active work involves a short initial transient during which the
particle is captured by the harmonic trap. Fig.\ \ref{fig:fig3}(b)
shows that this transient is characterized by a large value of the
initial position, $r(0)\sim\sqrt{\tau}$, which goes along with a large
value of the initial active force, $a(0)\sim\sqrt{\tau}$, in the same
direction since $r(0)$ and $a(0)$ are positively correlated by
Eq.\ (\ref{eq:staz_incond}). The contribution of these large values to
the active work is of order $-a(0)r(0)\sim\tau$ and negative because
the particle moves oppositely to the active force.  In conclusion, the
most likely way to realize the rare event $w\ll w_-$ is that an
initial transient provides a macroscopic fraction of the large
fluctuation, with the active force trying to push the particle out of
the harmonic trap unsuccessfully. Specularly, a large fluctuation
$w\gg w_+$ entails a final short transient during which the particle
escapes from the trap. In fact, Fig.\ \ref{fig:fig3}(c) proves that
there are large final values of the position and the active force,
$r(\tau)\sim\sqrt{\tau}$ and $a(\tau)\sim\sqrt{\tau}$, and that they
are in the same direction.  This time the active force successfully
pushes the particle out of the harmonic trap, so that the contribution
of these large values to the active work is positive and of order
$a(\tau)r(\tau)\sim\tau$.  None of the above transients is observed
when $w\approx\braket{w}$. According to Fig.\ \ref{fig:fig3}(d), the
distribution of the net displacement of the particle in a time
interval $\tau$ has only one peak at zero when $w\approx\braket{w}$
and two symmetric peaks due to final large values when $w\gg w_+$.

Finally, under the non-stationary initial condition with small
$\kappa$ and $\Pe$, where dynamical phase transitions do not occur at
stationarity, we observe singularities at both $w\ll w_-$ and $w\gg
w_+$. These singularities arise solely due to large values in the
initial condition, $r(0)\sim\sqrt{\tau}$ and $a(0)\sim\sqrt{\tau}$ as
shown by Fig.\ \ref{fig:fig3}(e), with the particle captured by the
harmonic trap providing a contribution of order $-a(0)r(0)\sim\tau$ to
the active work. The latter can be either negative or positive since
$r(0)$ and $a(0)$ are now uncorrelated.

The occurrence of a large deviation of the active work due to large
values of either $r(0)$ and $a(0)$ or $r(\tau)$ and $a(\tau)$ is
reminiscent of some big-jump phenomena observed in sums of independent
random variables \cite{jeon2000,zia2006,ARMENDARIZ2009,godrche2019}.
The latter works have understood that a fluctuation in the linear tail
of the rate function can be decomposed in two parts: many small
deviations in the same direction which sum up to the singular point,
and a big jump of a single variable summing to the actual value of the
fluctuation.  Basically, we find that a large fluctuation of the
active work is realized in a similar way through some localized big
jumps. At variance with sums of independent random variables where
summands are exchangeable, here the structure of the process imposes
that the big jumps localize at the initial or at the ending points of
the trajectories.  In fact, $r(t)$ and $a(t)$ are always positively
correlated.  Thus, suppose a big jump of $r(t)$ and $a(t)$ occurs at
an intermediate time $t$, with the particle escaping the potential up
to $t$ and generating a certain positive active work; afterwards, the
particle is bound to be recatched by the potential, and in doing so
generates a negative active work that cancels out the first
contribution.

In summary, we have characterized the active work large fluctuations
of an active Ornstein-Uhlenbeck particle under the action of a
harmonic potential. We have demonstrated that harmonic confinement can
induce dynamical phase transitions at sufficiently large active and
harmonic force parameters. Furthermore, we have provided an in-depth
understanding of the origin of these transitions in terms of phase
separation in trajectory space driven by big-jump mechanisms.  These
results can contribute to understand the origin of singularities in
active work rate functions in more complex systems of interacting
active Brownian particles. We argue that our approach can be extended
to the study of fluctuations in systems of several Ornstein-Uhlenbeck
particles coupled via elastic forces, like active polymers.

This work has been supported by the Italian Ministry of University and
Research via the project PRIN/2020 PFCXPE and by Apulia Region via the
project UNIBA044 of the research programme REFIN - Research for
Innovation.

\bibliographystyle{apsrev4-1}
\bibliography{Bibliography.bib}

\end{document}